\documentclass[%
 reprint,
 amsmath,amssymb,
 aps,
]{revtex4-2}

\usepackage{ragged2e}
\usepackage{float}
\usepackage{booktabs}
\usepackage{color}
\usepackage{graphicx}
\usepackage{dcolumn}
\usepackage{bm}

\begin{document}

\preprint{APS/123-QED}

\title{Sonogenetics is a novel antiarrhythmic treatment}%

\author{Yang Li$^{1,2}$}
\author{Xingang Wang$^{2}$} 
\author{Jianzhong Guo$^{2}$} 
\author{Yong Wang$^{3,4}$}
\author{Vladimir Zykov$^{3}$}
\author{Eberhard Bodenschatz$^{3,4,5,6}$}
\author{Xiang Gao$^{2,3,}$}
\email{gaoxiang.gnaixoag@gmail.com}

\affiliation{$^{1}$School of Science, Beijing University of Posts and Telecommunications, Beijing, 100876, China}
\affiliation{$^{2}$School of Physics and Information Technology, Shaanxi Normal University, Xi’an, 710062, China}
\affiliation{$^{3}$Laboratory for Fluid Physics, Pattern Formation and Biocomplexity, Max Planck Institute for Dynamics and Self-Organization, G{\"o}ttingen, 37077, Germany}
\affiliation{$^{4}$DZHK (German Center for Cardiovascular Research), Partner Site G{\"o}ttingen, G{\"o}ttingen, 37077, Germany}
\affiliation{$^{5}$Institute for Dyanmics of Complex Systems, University of G{\"o}ttingen,G{\"o}ttingen, 37075, Germany }
\affiliation{$^{6}$Laboratory of Atomic and Solid-State Physics and Sibley School of Mechanical and Aerospace Engineering, Cornell University, Ithaca, New York, 14853, United States of America}

\date{\today}

\begin{abstract}
Arrhythmia of the heart is a dangerous and potentially fatal condition. The current widely used treatment is the implantable cardioverter defibrillator (ICD), but it is invasive and affects the patient's quality of life. The sonogenetic treatment technique proposed here focuses transthoracic ultrasound on the heart, noninvasively stimulates endogenous stretch-activated Piezo1 ion channels on the focal region's cardiomyocyte plasma membrane, and restores normal heart rhythm. In contrast to anchoring the implanted ICD lead at a fixed position in the myocardium, the size and position of the ultrasound focal region can be selected dynamically by adjusting the signal phases of every piezoelectric chip on the wearable ultrasonic phased array, and it allows novel and efficient defibrillations. Based on the developed interdisciplinary electromechanical model of sonogenetic treatment, our analysis shows that the proposed ultrasound intensity and frequency will be safe and painless for humans and well below the limits established by the U.S. Food and Drug Administration.
\end{abstract}

\maketitle


\section{\label{sec:level1}Introduction}

Cardiac arrhythmias can lead to stroke, heart failure, or sudden death \cite{witkowski1998nature, holden1998nature, garfinkel2000pnas}. The established medical treatment of cardiac arrhythmias is the implantable cardioverter defibrillator (ICD) \cite{qu2014physicsreports}. The wires of the ICD are passed through a vein and terminate in electrodes that are anchored in the heart muscle. The ICD device itself is installed under the skin and measures the electrical activity of the heart. If cardiac arrhythmias are detected, the ICD delivers high-voltage electrical shocks to the heart muscle, resetting the electrophysiological system and restoring a normal heartbeat. This invasive and potentially very painful treatment carries significant risks, such as infection at the implant site and damage to blood vessels from the ICD wires. It adversely affects the patient's quality of life \cite{graham2005nature, walcott2003resuscitation, verma2004heartrhythm}. Therefore, cardiologists wish for less invasive antiarrhythmic treatments \cite{sankaranarayanan2013bjhm}.

\par Sonogenetics finds its first use in the non-invasive regulation of excitable cells, such as neurons \cite{ye2018nanoletters, rabut2020neuron, wang2020FrontiersinPhysiology} since the seminal work by Ibsen et al. in 2015 \cite{ibsen2015nc}. Neurons are excited or inhibited by the ionic current from stretch-activated ion channels \cite{chang1998science, lin2019nature, cox2016nc, lewis2015elife}. Their opening probability changes with the membrane tension, which can be non-invasively controlled by low-intensity transcranial ultrasound \cite{RN661, rabut2020}. One of the important stretch-activated ion channels is the human endogenous Piezo1 channel \cite{geNature2015, Mathias2015}.

\par Here, we propose sonogenetics to treat malignant electrical excitations of the heart, i.e., arrhythmia, without invasive electrical connections. We establish in-silico experiment procedures that can be accomplished with a wearable cardiac ultrasound device. We elaborate the electromechanical model of Piezo1 channel current under ultrasonic radiation pressure and study the electrical stimulation effect of Piezo1 channel currents to terminate the arrhythmia. 

Our results show that sonogenetics can effectively eliminate tachycardia and fibrillation, corresponding to the electrophysiological patterns of rotors and electrical turbulence, respectively \cite{ouyang2000prl}. The ultrasound we use in the simulations is within the U.S. Food and Drug Administration (FDA) safety standard \cite{FDA2019}. Our proof of concept demonstrates that sonogenetics is a prime candidate for the non-invasive and harmless treatment of malignant electrical excitations of a person's heart.

\section{\label{sec:level2}METHODS}

\begin{figure}[tp]
\centering
\includegraphics[width=\linewidth]{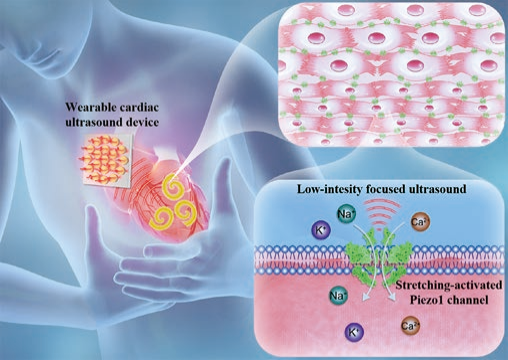}
\caption{An envisioned clinical application of sonogenetic treatment of arrhythmia. Left: To noninvasively treat arrhythmia, the wearable cardiac ultrasound phased array on the chest focuses the transthoracic ultrasound dynamically on any specified area of the heart. Top right: The stretching-activated Piezo1 channels endogenously express on the cardiomyocyte membrane and would upregulate in the hypertrophied after a heart failure. Bottom right: The ultrasonic radiation pressure activates the Piezo1 channels in the ultrasound-focused area. And Piezo1 channel currents excite action potential to eliminate arrhythmia.}
\label{fig1}
\end{figure}

\subsection*{Proof of concept}
Figure \ref{fig1} shows the procedures of our sonogenetic treatment of arrhythmia in an envisioned clinical application. The stretching-activated Piezo1 channels endogenously express on the human cardiomyocyte membrane and would upregulate in the hypertrophied after a heart failure. When a patient has rotors in the heart and suffers arrhythmia, the wearable cardiac ultrasound phased array on the chest adjusts each signal phase of piezoelectric chips. It focuses a low-intensity transthoracic ultrasound wave on any specified heart area. The ultrasonic radiation pressure opens the Piezo1 channels in the ultrasound-focused area. And Piezo1 channel currents excite action potential to eliminate arrhythmia. The detailed mechanism and success rate of our sonogenetic treatment are demonstrated in the Results section.

Each procedure of our sonogenetic treatment in the above envisioned clinical application has been implemented respectively: 1. Wearable cardiac ultrasound phased arrays \cite{wang2018, WangScience2022, hu2023} focus ultrasound through ribs \cite{qiao2013, greef2015, chao2015, wu2017tbs, cao2020, Wang2021}. The focused ultrasound field can be in a dot, line, planar, or arbitrary shape like a bird \cite{foresti2013, melde2016, marzo2019pnas}. 2. The Piezo1 channel we investigated is a bona fide mechanosensitive cation channel that utilizes its unique three-bladed, propeller-like architecture to most effectively convert distinct forms of mechanical stimuli such as low-intensity and harmless transthoracic ultrasound into Ca$^{2+}$, Na$^+$, K$^+$, etc. signaling \cite{cox2017, qiu2019iscience, Ridone2019, yooNC2022}. 3. Piezo1 is abundant in the human heart and is upregulated folds in failure hearts \cite{Liang2017, jiang2021nc, Ploeg2021, Bartoli2022, Yu2022}. Therefore, no need for virus transfection as used in sonogenetics for neuromodulation.

\subsection*{In-silico experiment}
To demonstrate the mechanism of sonogenetic arrhythmic control, we establish in-silico experiments. We studied a slice of cardiac tissue placed in a culture container and immersed in a culture solution. Low-intensity focused ultrasound propagating from above forms dot-, line-, or ring-shaped spatial distributions of the ultrasonic intensity $I_{0}$ on the cardiac tissue. According to previous sonogenetics experiments \cite{menz2019JournalofNeuroscience}, $I_{0}$ controls the spatial distribution of the ultrasonic radiation pressure $\varGamma$ (see Appendix A for the derivation of ultrasound radiation pressure in our in-silico experiments):
\begin{equation} \label{eq:1}
\begin{split}
\varGamma c=2 I_0,
\end{split}
\end{equation}
where the velocity of ultrasound $c$ in the cardiac tissue is $1561.3 \rm{m/s}$. $\varGamma$ applied on the cardiac slice presses the myocardium and opens Piezo1 channels. 
\par To model the transition of Piezo1 channel's states under different pressures, Lewis et al. proposed a four-state model based on experimental data \cite{lewis2017cellreports}. We assume that the ultrasound radiation pressure $\varGamma$ is the pressure on the Piezo1 channels in the cardiomyocyte membranes. We modify their model to obtain a four-state model for the Piezo1 channel controlled by ultrasonic radiation pressure $\varGamma$:
\begin{equation} \label{eq:2}
\begin{split}
\frac {dO_{\rm{Piezo1}}} {dt}=a(\varGamma) C + I_1 d + h I_2 - (b+c+g) O_{\rm{Piezo1}},   
\end{split}
\end{equation}
where $O_{Piezo1}$, $C$, $I_1$, and $I_2$ are the probabilities that the Piezo1 channel is in open, closed, and two inactivation states, respectively. The transition rates between these four states are represented by $a(\varGamma)$, $b$, $c$, $d$, $e(\varGamma)$, $f$, $g$, and $h$, where $a(\varGamma)$ and $e(\varGamma)$ are related to the ultrasonic radiation pressure $\varGamma$ (see Fig. \ref{A1}). The detailed description and equations of the four-state model for the Piezo1 channel are shown in Appendix B.

\par Then, we use the stochastic modeling to describe the current equation of Piezo1 channels: 
\begin{equation} \label{eq:3}
\begin{split}
I_{\rm{Piezo1}}=\bar{g}_{\rm{Piezo1}}N_{\rm{Piezo1}}O_{\rm{Piezo1}}(\varGamma)(V-E_{\rm{Piezo1}}),
\end{split}
\end{equation}
where $\overline{g}_{\rm{Piezo1}}$ and $E_{\rm{Piezo1}}$ are the maximal conductance of a Piezo1 channel and the reversal potential of $27.3 \rm{pS}$ and $8.8 \rm{mV}$ measured in the experiment \cite{saotome2018nature}; $N_{\rm{Piezo1}}$ is the Piezo1 channel's density, which indicates the total number of Piezo1 channels per $\rm{cm^2}$; $O_{\rm{Piezo1}}$ is the open probability of Piezo1 channels; $V$ is the membrane potential.

\par We propose to add Piezo1 channels' current to a normal cardiac model to obtain a cardiac model regulated by sonogenetics. Here we use the Fenton-Karma three-variable model \cite{fenton1998chaos, fenton2002chaos} (see details in Appendix C) to simulate the electrophysiological activities of human tissue. This model consists of three variables: the membrane potential $V$, a fast ionic gate $v$, and a slow ionic gate $w$. The three variables are used to produce three independent phenomenological currents: a fast inward inactivation current $I_{fi}$, a slow time-independent rectifying outward current $I_{so}$, and a slow inward inactivation current $I_{si}$. We add Piezo1 channels' current to this model, and the membrane potential equation is as follows:
\begin{equation} \label{eq:4}
\begin{split}
\partial_{t}V= & \nabla (D_V \nabla V) \\
 & - \frac{[I_{fi}(V,v)+I_{so}(V)+I_{si}(V,w) + I_{\rm{Piezo1}}(V,\varGamma)]}{C_m}, 
\end{split}
\end{equation}
where $D_V=0.001 \rm{cm^2/ms}$ is the diffusion constant, and $C_m= 1 \mu \rm{F/cm^2}$ is the membrane capacitance. For 2D simulation, no-flux boundary conditions are used. Time evolutions are calculated using an explicit Euler method. The diffusion part in the cardiac model is calculated by the five-point stencil method. The time and space steps are $0.05 \rm{ms}$ and $0.025 \rm{cm}$.

\par At a given open probability of Piezo1 channels, their channel density $N_{\rm{Piezo1}}$ needs to exceed some threshold so that Piezo1 channel currents become large enough to stimulate a propagating action potential at the resting state of cardiac tissue (see Fig. \ref{A2}). Since $\varGamma$ controls the open probability of Piezo1 channels, the threshold of $N_{\rm{Piezo1}}$ is also controlled by $\varGamma$, as illustrated by the blue line and circular symbols in Fig. \ref{A3}.

\subsection*{FDA guidance for safe cardiac ultrasound}
According to an FDA guidance \cite{FDA2019}, the spatial-peak temporal-average intensity ($ISPTA$) of medical pulsed ultrasound applied to the heart should not exceed $430 \rm{mW/cm^{2}}$. $ISPTA=DI_0/1 \rm{s}$, where $D$ and $I_0$ are the ultrasonic duration and intensity, respectively. In our in-silico experiments, $2I_{0}=\varGamma c$ (see Eq. \ref{eq:1} in Methods). Thus, the ultrasonic radiation pressure $\varGamma$ and duration $D$ are limited to $\varGamma D \leq \frac {2 \times 430 \rm{mW/cm^{2}}}{c}$. The red line and square symbols in Fig. \ref{A3} illustrate the upper limitation between the ultrasonic radiation pressure $\varGamma$ and ultrasonic duration $D$ for safe sonogenetic arrhythmia control.

\par Except for guidance to the $ISPTA$, FDA also uses the mechanical index ($MI$) and the thermal index ($TI$) to require the safety of ultrasound regarding cavitation and thermal effects, respectively \cite{FDA2019, BMUS2009, kollmann2013UltraschallinderMedizin, Nowicki2020}. The mechanical index $MI=p/\sqrt{f}$, where $p$ is the peak negative pressure in $\rm{MPa}$ and $f$ is the ultrasonic frequency in $\rm{MHz}$. To satisfy FDA guidance for $MI \leq 1.9$, considering $p=\sqrt{2\rho c I_0}$ (the cardiac tissue density $\rho=1081 \rm{kg/m^3}$) and $2I_{0}=\varGamma c$, the minimum value of ultrasonic frequency $f_{min}^{MI}=0.35\varGamma[\rm{mmHg}]/1.9^2$. The thermal index $TI$ for soft tissues is defined as $A\,f\,ISPTA / 210 $, where $A$ is the ultrasound focus area in $[\rm{cm^2}]$ and $ISPTA$ in $[\rm{mW/cm^{2}}]$. To satisfy FDA guidance for $TI \leq 6$, the maximum value of ultrasonic frequency $f_{max}^{TI}= 2.93/A[\rm{cm^2}]$. Thus, $\varGamma$ and $A$ control the existence and range of the ultrasonic frequency $f$ for safe sonogenetic arrhythmia control.

\par In the following in-silico experiments for sonogenetic arrhythmia control under FDA guidance about $ISPTA$, $MI$, and $TI$, we use the ultrasonic radiation pressure $\varGamma=41.35 \rm{mmHg}$, the ultrasonic duration $D=1 \rm{ms}$, Piezo1 channels' density $N_{\rm{Piezo1}}=1 \times 10^7 \rm{cm^{-2}}$, unless otherwise. The ultrasound phased array controls the ultrasound focus area $A$ dynamically. Based on different spatial-temporal distributions of $A$, we demonstrate five sonogenetic mechanisms for safely treating cardiac arrhythmia.

\section{\label{sec:level3}Results}

\begin{figure*}
\centering
\includegraphics[width=0.8\linewidth]{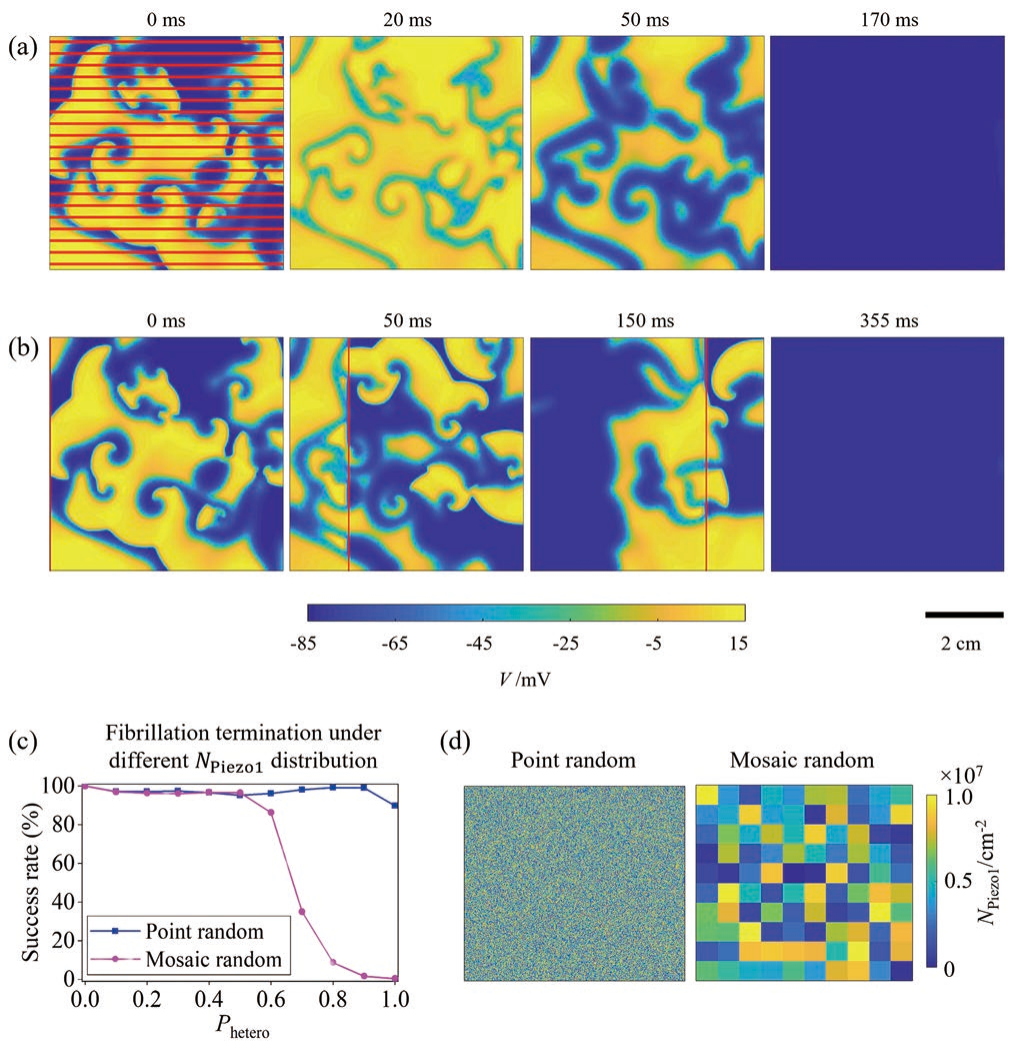}
\caption{Sonogenetic treatment of fibrillation. (a) Fibrillation termination by global ultrasound. Global ultrasound (indicated by the red shadow at $0 \rm{ms}$) is applied to a homogeneous cardiac tissue. Transmembrane potential $V$ in fibrillation is first turbulent ($0 \rm{ms}$), then synchronized to be excited ($20 \rm{ms}$), refractory ($50 \rm{ms}$), and reset to the resting state ($170 \rm{ms}$) by the ultrasound-stimulated Piezo1 channels' current. (b) Fibrillation termination by focused ultrasound stripe sweeping. Transmembrane potential $V$ in fibrillation is first turbulent ($0 \rm{ms}$). A focused ultrasound stripe width of $0.05 \rm{cm}$ (indicated by red lines) sweeps from left to right at a moving velocity $0.05 \rm{cm/ms}$ ($50$ and $150 \rm{ms}$). After swept, regions behind the stripe are excited by the focused ultrasound-stimulated Piezo1 channels' current, become refractory and reset to the resting state ($355 \rm{ms}$) sequentially. (c) Effect of heterogeneous upregulation of Piezo1 channels on the success rate of fibrillation termination by focused ultrasound stripe sweeping. $P_{\rm{hetero}}$ is the heterogeneity extent deviated from homogeneous Piezo1 channels' upregulation. (d) The heterogeneities at cell and tissue levels are simulated by the spatial distribution of Piezo1 channels' density $N_{\rm{Piezo1}}(x,y)$ in point and mosaic random, respectively. Tissue size is $10 \rm{cm} \times 10 \rm{cm}$. For other parameters in the above simulations, ultrasonic radiation pressure $\varGamma = 41.35 \rm{mmHg}$, ultrasonic duration $D=1 \rm{ms}$, and Piezo1 channels' density $N_{\rm{Piezo1}} = 1 \times 10^{7} \rm{cm^{-2}}$. See Movie S1 and S2 for animations corresponding to panels (a) and (b) in Supplemental Material \cite{supplementalmaterial}.}
\label{fig2}
\end{figure*}

\subsection*{Fibrillation termination by global ultrasound}
We first demonstrate that global ultrasound stimulation can eliminate fibrillation, which is self-sustained electrical turbulence on the cardiac tissue. The global ultrasound synchronizes the excited and refractory state of the whole tissue by globally stimulating the current of Piezo1 channels. It is similar to the high-energy electric shock in ICD. As shown in Fig. \ref{fig2}(a) (and Movie S1 in Supplemental Material \cite{supplementalmaterial}), at $t=0 \rm{ms}$ a global ultrasound pulse is applied on cardiac tissue in electrical turbulence. At $t=20 \rm{ms}$, the whole tissue is excited by the ultrasound-stimulated Piezo1 channels' current. At $t=50 \rm{ms}$, the tissue becomes refractory, which blocks electrical excitation propagation. At $t=170 \rm{ms}$, the cardiac tissue is reset to the resting state, and electrical turbulence is eliminated.

\par Our global ultrasound termination of fibrillation complies with the FDA guidance about $ISPTA$. But $f_{min}^{MI}=4.01 \rm{MHz}$ is larger than $f_{max}^{TI}=0.03 \rm{MHz}$. This means there is no range of ultrasonic frequency $f$ that can meet both the FDA guidance about $MI$ and $TI$. The reason is that the ultrasonic focus area $A$ is too large. Therefore, we need new approaches that reduce $A$ and thus increase $f_{max}^{TI}$ over $f_{min}^{MI}$ to have a suitable ultrasonic frequency range under all FDA guidances.

\subsection*{Fibrillation termination by focused ultrasound stripe sweeping}

We propose using local ultrasonic stimulation to comply with all FDA guidance by reducing the ultrasonic focus area $A$. Adjusting the signal phases of every piezoelectric chip on the ultrasonic phased array allows a stripe of locally focused ultrasound to sweep across the cardiac tissue. As shown in Fig. \ref{fig2}(b), at $t=0 \rm{ms}$, the transmembrane potential $V$ of the cardiac tissue is in the electrical turbulence of fibrillation, and a focused ultrasound stripe of width $0.05 \rm{cm}$ starts from the left edge. At $50 \rm{ms}$ and $150 \rm{ms}$, the stripe sweeps from left to right at a moving velocity $0.05 \rm{cm/ms}$. Regions swept by the stripe are sequentially excited by the focused ultrasound-stimulated Piezo1 channels' current, then become refractory zones that block turbulence propagation and reset to the resting state. At $t=355 \rm{ms}$, the whole cardiac tissue returns to the resting state, and the electrical turbulence of fibrillation is eliminated.

\par The effective ultrasonic duration $D$ of every node on the cardiac tissue is the ratio of the focused ultrasound stripe width over its moving velocity, i.e., $D= 1 \rm{ms}$. Thus, the sweeping focused ultrasound stripe termination of fibrillation complies with the FDA guidance about $ISPTA$. The ultrasonic focus area $A$ is the area of the stripe, which is $0.05 \rm{cm} \times 10 \rm{cm}= 0.5 \rm{cm^2}$. Thus, $f_{min}^{MI}= 4.01 \rm{MHz}$ and $f_{max}^{TI}= 5.86 \rm{MHz}$. This means an ultrasonic frequency $f$ between $4.01 \rm{MHz}$ and $5.86 \rm{MHz}$ can meet both FDA guidance about $MI$ and $TI$. Except for the above parameters, we demonstrate a phase diagram of the parameter range for successful and safe fibrillation termination by focused ultrasound stripe sweeping in Fig. \ref{A4}. 

\par In the above simulations with different initial conditions of electrical turbulence, the success rate of fibrillation termination is $100\%$ at homogeneous Piezo1 channels' density $N_{\rm{Piezo1}} = 1 \times 10^{7} \rm{cm^{-2}}$. For sonogenetic arrhythmia control in the real heart, we studied the effect of heterogeneous upregulation of Piezo1 channels on the success rate. As shown in the subfigure of Fig. \ref{fig2}(c), the heterogeneities at cell and tissue levels are simulated by the spatial distribution of Piezo1 channels' density $N_{\rm{Piezo1}}(x,y)$ in point and mosaic random, respectively. $N_{\rm{Piezo1}}(x,y)=(1-\xi(x,y) P_{\rm{hetero}}) \times 10^{7} \rm{cm^{-2}} $, where $\xi(x,y)$ are point- or mosaic-distributed random numbers taking values between $0$ and $1$, and $P_{\rm{hetero}}$ controls the heterogeneity extent deviated from homogeneous Piezo1 channels' upregulation. As shown in Fig. \ref{fig2}(c), the Piezo1 channels' upregulation heterogeneity at the cell level (the black line) barely changes the high success rate of fibrillation termination by focused ultrasound stripe sweeping. But for the heterogeneity at the tissue level (the red line), the success rate drops when $P_{\rm{hetero}}$ approaches to 1.

\par The different results between cell- and tissue-level heterogeneities are because the Piezo1 channels' density at each node needs to be larger than $4.1\times10^6 \rm{cm^{-2}}$ so that the ultrasonic radiation pressure used in the above simulations can excite them, as illustrated in Fig. \ref{A2}. When $P_{\rm{hetero}}=1$, $41\%$ of nodes is unexcitable by a focused ultrasound stripe. In the case of cell-level heterogeneity, these unexcitable nodes are distributed point randomly among the nodes excitable by ultrasound and would be excited soon by neighboring excited nodes. Thus swept by the focused ultrasound stripe, regions consisting of excitable and unexcitable nodes are still sequentially excited, refractory, and reset to the resting state. The mechanism of terminating fibrillation still works. But for the case of tissue-level heterogeneity, unexcitable nodes are in random mosaic distribution. They are clustered and unsusceptible to excitation from neighboring excited nodes. And parts of regions swept by focused ultrasound stripe are not sequentially excited, refractory, and rest to the resting state, as shown in Fig. \ref{A5}. Therefore, the success rate drops with the increase of tissue-level heterogeneity extent of Piezo1 channels upregulation.

\par The novel of fibrillation termination by focused ultrasound stripe sweeping is that, unlike ICD electrodes' fixed connection to the endocardium, spatial-temporal selective stimulation of ultrasonic phased arrays can change the focus position in real-time and dynamically stimulate electrical excitations to sweep turbulence away.

\begin{figure*}
\centering
\includegraphics[width=0.8\linewidth]{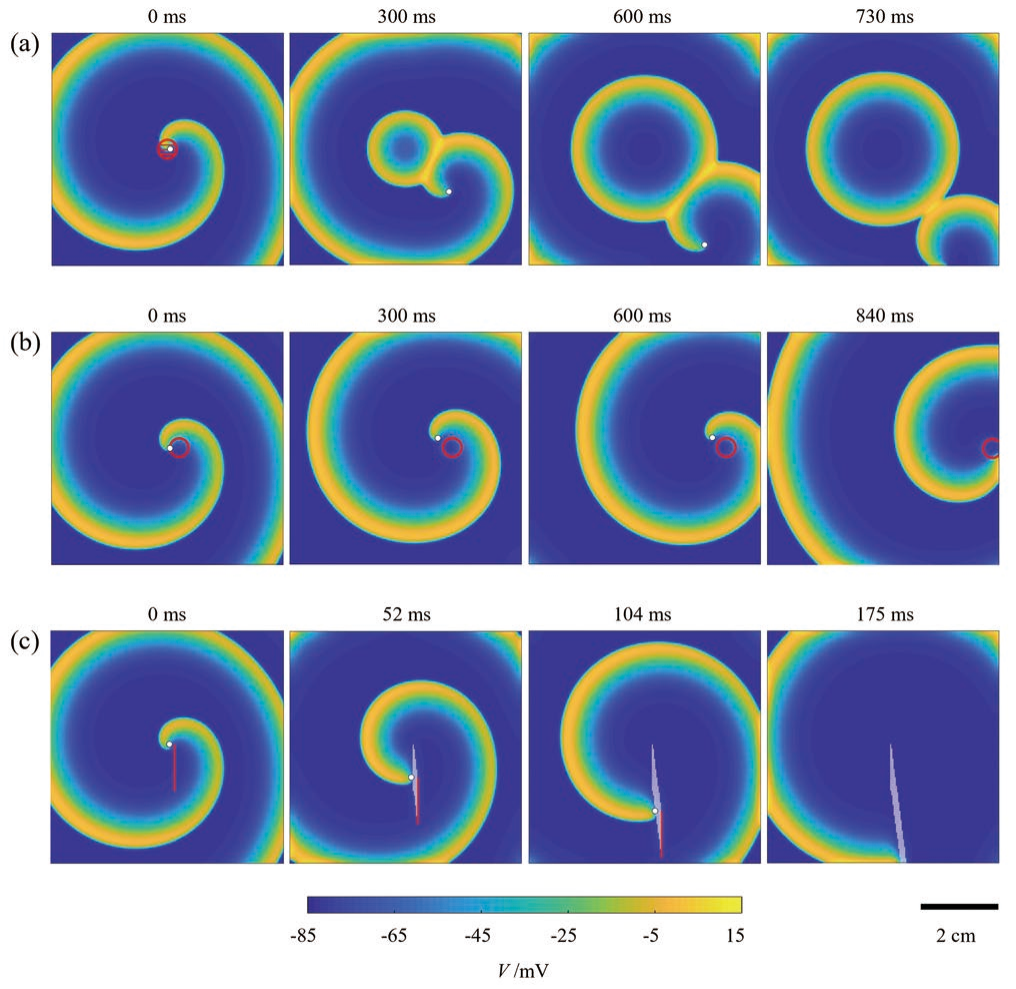}
\caption{Sonogenetic treatment of tachycardia. (a) Overriding rotor by focused ultrasound pulses. At $t=0 \rm{ms}$, the transmembrane potential $V$ starts as a rotor, which rotates at a $6.3 \rm{Hz}$ frequency. The pulsed ultrasound is focused on a circular area (indicated by a red shadow) with a radius of $0.4 \rm{cm}$ in the center of the tissue. The ultrasonic pulse frequency is $8 \rm{Hz}$, and each pulse lasts $0.1 \rm{ms}$. The rotor is overridden by high-frequency target waves stimulated by focused ultrasound at $t= 300$ and $600 \rm{ms}$ and driven out of the tissue at $t= 730 \rm{ms}$. The white dots indicate rotor tips. The ultrasonic radiation pressure $\varGamma = 51.68 \rm{mmHg}$. (b) Pulling rotor away by focused ultrasound ring. Transmembrane potential $V$ starts as a rotor at $t=0 \rm{ms}$. The pulsed ultrasound is focused on a ring-shaped region (indicated by the red ring). The ring width is $0.025 \rm{cm}$, the radius of the inner circle is $0.4 \rm{cm}$, and the moving speed of the ring is $0.005 \rm{cm/ms}$. The ultrasonic pulse frequency is $200 \rm{Hz}$, and each pulse lasts $0.5 \rm{ms}$. The focused ultrasound creates a ring-shaped refractory zone that pins the rotor. The moving ultrasound ring pulls the rotor towards the tissue boundary at $t= 300$ and $600 \rm{ms}$ and away at $t= 840 \rm{ms}$. The ultrasonic radiation pressure $\varGamma = 41.35 \rm{mmHg}$. (c) Leading rotor out by focused ultrasound fast-channel staircase. The transmembrane potential $V$ starts as a rotor at $t=0 \rm{ms}$. The ultrasound is focused on a rectangle area with a $2 \rm{cm}$ length and $0.05 \rm{cm}$ width (indicated by the red line) and creates a refractory zone. For each ultrasound pulse at $13 \rm{ms}$ intervals, the ultrasound-focused area moves bottom right next to the previously focused area. The white shadows indicate once-focused rectangle areas at previous pulses. They form a stair-case refractory zone as a fast channel that leads the rotor towards the tissue boundary (at $t= 52$ and $104 \rm{ms}$) and away (at $t= 175 \rm{ms}$). The ultrasonic radiation pressure $\varGamma= 41.35 \rm{mmHg}$. For other parameters in the above simulations, tissue size is $10 \rm{cm} \times 10 \rm{cm}$, and Piezo1 channels density $N_{\rm{Piezo1}} = 1 \times 10^{7} \rm{cm^{-2}}$. See Movie S3-S5 for animations in Supplemental Material \cite{supplementalmaterial}.}
\label{fig3}
\end{figure*}

\subsection*{Overriding rotor by focused ultrasound pulses}
Tachycardia clinically corresponds to rotors and may turn into lethal fibrillation. In hearts as excitable media, high-frequency waves can drive the low-frequency out of the medium boundary \cite{cao2007chaos}. Accordingly, ICD electrodes pace the heart with high-frequency excitation waves and override low-frequency rotors, which is called anti-tachycardia pacing \cite{echt1985circulation}. We propose to use focused ultrasound pulses instead of ICD electrodes to generate high-frequency target waves because it has the following advantages. 1. According to the eikonal relation of the excitable medium \cite{xie1999pre}, a large enough radius stimulated region is needed to excite target waves. The electrode of the ICD needs to be a large loop but may detach as the heart beats. The focal size of the ultrasound can enlarge accordingly. 2. The position of electrodes is fixed after implantation, may be away from the rotor tip, and thus leads to less-effective anti-tachycardia pacing. The focal position of ultrasound can be dynamically placed near the rotor tip. 3. The electrode of ICD is attached to the endocardium, and excitations can only be stimulated at the surface of the heart wall. Thus, it is hard to terminate scroll waves in the myocardium. Focused ultrasound can noninvasively stimulate excitations anywhere in the heart wall.

\par Figure \ref{fig3}(a) (and Movie S3 in Supplemental Material \cite{supplementalmaterial}) shows how focused ultrasound stimulates high-frequency target waves ($8 \rm{Hz}$) to override the low-frequency rotor ($6.3 \rm{Hz}$) out of the tissue boundary. We use the Jacobian-determinant method \cite{li2018pre} to mark the position of the tip of the rotor (white dot in Fig. \ref{fig3}(a)). If the tip leaves the tissue, it can be determined that the rotor will disappear. At $t=0 \rm{ms}$, the focused ultrasound is pulsed on a circular area (red shadow) to generate target waves. Then, the rotor is gradually pushed to the boundary by the ultrasonic target wave, and the tip also moves to the boundary (as shown in the $300$ and $600 \rm{ms}$ snapshots in Fig. \ref{fig3}(a)). At $t=730 \rm{ms}$, the tip of the rotor leaves the tissue, and then the rotor gradually disappears. After the rotor is eliminated, we stop the ultrasonic target wave stimulation, and the cardiac tissue returns to its excitable state. In this case, according to the guidance for the $ISPTA$, we use $D=0.1\times8=0.8 \rm{ms}$ and $\varGamma = 51.68 \rm{mmHg}$. Then, we calculate that $f_{min}^{MI}=5.01 \rm{MHz}$ and $f_{max}^{TI}=5.83 \rm{MHz}$. 

\par Ultrasonic target waves can effectively eliminate rotors. However, the frequency of target waves will not continuously respond to the increasing frequency of the ultrasonic pulses. The frequency of ultrasonic target waves has an upper limit, at which phase-locking phenomena change (see Fig. \ref{A6}). Therefore, the approach of overriding by ultrasonic series is only applicable for slow rotors.

\subsection*{Pulling rotor away by focused ultrasound ring}
Besides using electrical excitations stimulated by focused ultrasound, dynamically creating refractory zones by shorter-duration focused ultrasound can also eliminate rotors. By adjusting the phase and duration of each ultrasonic transducer among the array, we can create a ring-shaped refractory zone in the cardiac tissue. As shown in Fig. \ref{fig3}(b) (and also Movie S4 in Supplemental Material \cite{supplementalmaterial}), at $t=0 \rm{ms}$, ultrasound short pulses continuously generate a ring-shaped refractory zone (red ring) near the rotor tip (white dot). The rotor cannot pass through this refractory zone and will revolve around it, which we call pinning. Then, we slowly move the ultrasonic ring toward the tissue boundary and pull the pinned rotor away. At $t=300$, and $600 \rm{ms}$, the pinned rotor is pulled to the right until its tip is pulled out of the tissue boundary at $t=840 \rm{ms}$. 

\par To meet the FDA guidance about the $ISPTA$, we set the ultrasonic pulses to stimulate once every $5 \rm{ms}$, each lasting $0.5 \rm{ms}$. Therefore ultrasound stimulates each node in the tissue for $1 \rm{ms}$ at most. In addition, due to the small area of the focused ultrasound ring, the ultrasonic frequency range is wide  ($f_{min}^{MI}=4.01 \rm{MHz}$ and $f_{max}^{TI}=45.22 \rm{MHz}$). The phase diagram of the ultrasonic ring's moving velocity and direction that can successfully eliminate this rotor under FDA guidance is shown in Fig. \ref{A7}.

\subsection*{Leading rotor out by focused ultrasound fast-channel staircase}
To eliminate the rotor faster, focused ultrasound can also create a staircase refractory zone as a “fast channel" that leads the rotor out. As shown in Fig. \ref{fig3}(c), at $t=0 \rm{ms}$, a short ultrasound pulse (red line) creates a rectangle refractory zone as the first staircase. The rotor cannot pass through it and will propagate along the staircase. As the rotor propagates to the end of the previous staircase, another staircase refractory zone would be created next to it. All staircase refractory zone (white lines) forms a fast channel that leads the rotor towards the tissue boundary (at $t= 52$ and $104 \rm{ms}$) and away (at $t= 175 \rm{ms}$). The $ISPTA$ meets the guidance of the FDA, and the ultrasonic frequency range is from $f_{min}^{MI}=4.01 \rm{MHz}$ to $f_{max}^{TI}=29.30 \rm{MHz}$.

\par The success of the above two rotor terminations relies on the ability to dynamically create refractory zones by short-pulses focused ultrasound, which ICD electrodes cannot.

\section{\label{sec:level4}Discussion}

\subsection*{Advantages over other defibrillation methods}
Compared with invasive ICDs, which require wires to be implanted through the vein and fixed electrodes on the endocardium for delivery of electrical shocks, our sonogenetics-based method could place the ultrasonic phased array outside the patient's chest and adjust phases of the array to focus ultrasonic waves on selective areas for effective defibrillation. Compared with optogenetic defibrillation \cite{Arrenberg2010science, Bruegmann2010nature, Bingen2011Cardiovascularresearch, Burton2015nature, Entcheva2016Journalofphysiology, Majumder2018elife, Majumder2022PRA}, which could only illuminate the surface of the heart wall, the ultrasound can penetrate it and control the cardiac excitation transmurally. Compared with previous ultrasonic defibrillation \cite{smailys1981Resuscitation, Echt2007, Kohut2010}, our sonogenetic treatment uses the wearable ultrasonic phased array to focus a low-intensity ultrasound on a small heart area and controls the endogenous Piezo1 channels on the human cardiomyocyte membrane to terminates arrhythmia harmlessly under FDA safety standards. Recently, a cooling-based method has been proposed \cite{Majumder2021fphys}, but its clinical implementation remains challenging.

\subsection*{Long-term side effects of Piezo1 upregulation}
Recently, it was reported that the upregulation of Piezo1 channels as a response to pressure overload after heart failure may initiate cardiac fibrosis and hypertrophy in adult mice \cite{Bartoli2022, Yu2022}. To block the signaling pathway of fibrosis and hypertrophy, Piezo1 is a potential therapeutic target. Sonogenetics is a promising physiological control because it allows noninvasive and spatiotemporal activation and deactivation of Piezo1 with low-intensity focused ultrasound from a wearable cardiac ultrasound phased array. 

\subsection*{Potential applications for atrial/ventricular arrhythmia}
Although our in-silico experiments are on a two-dimensional cardiac tissue, one of the advantages of sonogenetic arrhythmia control is the penetration of ultrasound into the three-dimensional heart wall. Thus, it can noninvasively and dynamically eliminate malignant electrical waves and turbulence during atrial/ventricular arrhythmia. The difficulty of focusing ultrasound through the heterogeneous rib cage and heart has been solved by the same technologies used in transcranial ultrasound therapy \cite{maimbourg2018, gambin2019, bancel2021, qiu2021}. For example, a high-intensity transthoracic ultrasound is focused on a porcine aortic valve in-vivo to noninvasively treat calcific aortic stenosis \cite{Messas2020}. Clinical ultrasound systems, as well as wearable ultrasound devices, operate at a frequency range between $2$ and $10 \rm{MHz}$ \cite{carovac_smajlovic_junuzovic_2011, wang_kesteven_huttner_feneley_fatkin_2018, Wang2021, WangScience2022}, precisely the ultrasound frequency we used. Therefore, focusing ultrasound at frequencies we used through the heterogeneous rib cage and heart is feasible.

\par The successes and advantages over the ICD of our sonogenetic treatment of atrial arrhythmia are well demonstrated in our results. As for ventricular arrhythmia, the ultrasonic phased array can extend the sweeping focused ultrasound stripe to a focused ultrasound planar in a three-dimensional ventricle sweeping from the atrioventricular septum to the apex. The sweeping stripe terminating turbulence mechanism also extends to a three-dimensional case. The equations of $MI$ and $TI$ need to be reconsidered in three-dimensional cases in future studies. As for extending the mechanisms of the ultrasound ring and fast-channel staircase to the three-dimensional ventricular anti-tachycardia, reconstruction of the electrical scroll waves from other clinical diagnostic data is needed. Combining three-dimensional mechanical scroll waves echoed by high-resolution four-dimensional ultrasound-based strain imaging \cite{Christoph2018} with inverse mechano-electrical reconstruction by machine learning is promising \cite{ChristophChaos2020, HerzogFrontiers2021}.

\par The filaments of electrical scroll waves may be complex and need well-designed refractory zones to lead them away, rather than an ultrasound ring or fast-channel staircase. But the mechanism of dynamically creating refractory zones by lower-intensity focused ultrasound minimizes the risk of triggering extra excitations. Leading rotors away are similar to two- or three-dimensional defibrillations in the “teleportation" mechanism \cite{detal_kaboudian_fenton_2022}. And the ability of spatial-temporal stimulation by noninvasive focused ultrasound can be the prerequisite technology for clinical applications of the theoretical mechanism of teleportation.

\section{\label{sec:level5}Conclusions}
Based on our first interdisciplinary electromechanical model of the endogenous Piezo1 channel under non-invasive focused ultrasound, we demonstrate that sonogenetics is an excellent candidate for harmless antiarrhythmic treatment. A combination of existing experimental methods can accomplish our numerical results. The potential clinical application can be a wearable cardiac ultrasound phased array for noninvasive diagnosis and treatment of arrhythmias.

\begin{acknowledgements}
This work is supported by the National Natural Science Foundation of China (Program No. 11727813), Natural Science Basic Research Program of Shaanxi (Program No.2019JQ-163 and NO.2019JQ-810), the Fundamental Research Funds for the Central Universities No. GK202103011, the Max Planck Society, the German Center for Cardiovascular Research, and BMBF through the project IndiHEART. We thank Dr. Amanda H. Lewis from Duke University Medical Center for sharing experimental data about the Piezo1 channel, and Prof. Bingwei Li from Hangzhou Normal University, Prof. Haihong Li, Prof. Junzhong Yang, and Prof. Qionglin Dai from Beijing University of Posts and Telecommunications, Prof. Hong Zhang from Zhejiang University, Prof. Huaguang Gu from Tongji University, Prof. Mengjiao Chen from Leshan Normal University, Prof. Siyuan Zhang from Xi’an Jiaotong University, and Prof. Wei Ren from Shaanxi Normal University for helpful discussions.
\end{acknowledgements}

\section*{APPENDIX A: Derivation of ultrasound radiation pressure in our in-silico experiments}
\par In ultrasonic propagation, radiation pressure will be produced due to the difference in spatial energy density of the sound field. In our in-silico experiment, the ultrasonic phased arrays and cardiac tissue are immersed in a liquid culture medium. The direction of the ultrasonic incident wave is perpendicular to the surface of cardiac tissue. Ultrasound first propagates through the culture medium and then passes through the cardiac tissue to the bottom of the culture container. The attenuation coefficient of the tissue culture medium is very small, and the thickness of cardiac tissue can be ignored. So, ultrasound will not lose energy because of absorption attenuation. 

However, the reflection and transmission of ultrasound through the interface will produce energy differences. The ultrasound passes through two interfaces during its propagation: culture medium-cardiac tissue interface and cardiac tissue-culture container interface. The ultrasonic radiation pressure per unit area on the interface is:
\begin{equation} 
\varGamma=\langle E_i\rangle+R\langle E_i\rangle-T\langle E_i\rangle, \tag{A1}
\end{equation}
\begin{equation}
\langle E_i\rangle=\frac{I_{0}\cos{(2\pi x/ \lambda})}{c}, \tag{A2}
\end{equation}
where $\langle E_i\rangle$ is the average energy per unit volume of the incident wave. $I_{0}$ is the ultrasonic intensity; $x$ is the distance from ultrasonic phased arrays to the cardiac tissue; $\lambda$ is the ultrasonic wavelength, depending on the ultrasonic frequency; $c$ is the velocity of ultrasound in the medium, which is 1561.3 \rm{m/s} in cardiac tissue \cite{tissuedata}. $R$ and $T$ are the reflection coefficient and transmission coefficient, which can be calculated as \cite{naorjne2016}:
\begin{equation}
R=(\frac{Z_{out}-Z_{in}}{Z_{in}+Z_{out}})^2, \tag{A3}
\end{equation}
\begin{equation}
T=\frac{4Z_{in}Z_{out}}{(Z_{in}+Z_{out})^2}, \tag{A4}
\end{equation}
where $Z_{in}$ and $Z_{out}$ are the acoustic impedances of the incident side and the other side of the interface, calculated as the density $\rho$ multiplied by sound velocity $c$ of the medium. The acoustic impedances of tissue culture fluid and cardiac tissue are approximately equal \cite{tissuedata}. However, there is a great difference in the acoustic impedances between cardiac tissue and the culture container. Therefore, ultrasound approximately completely transmits  at the culture medium-cardiac tissue interface ($R=0$ and $T=1$) with no energy difference and approximately completely reflects at the cardiac tissue-culture container interface ($R= 1$ and $T= 0$) with energy difference. The ultrasonic radiation pressure at the culture medium-cardiac tissue interface is:
\begin{equation}
\varGamma=2\langle E_i\rangle=\frac{2I_{0}\cos{(2\pi x/ \lambda})}{c}. \tag{A5}
\end{equation}
To maximize the radiation pressure, we can adjust $x$ to an integer multiple of $\lambda$. Then, the equation becomes:
\begin{equation}
\varGamma c=2I_{0}. \tag{A6}
\end{equation}

\renewcommand{\thefigure}{A\arabic{figure}}
\setcounter{figure}{0}

\begin{figure}[t]
\centering
\includegraphics[width=\linewidth]{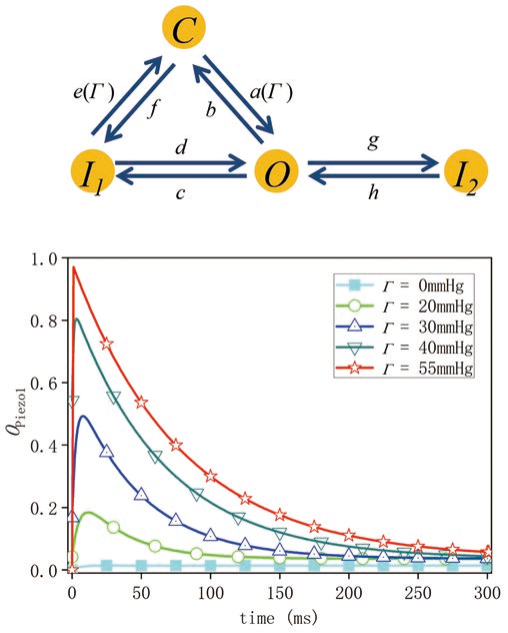}
\caption{Four-state model for the Piezo1 channel. (Top) Diagram for the Piezo1 channel's open ($O$), closed ($C$), and two inactivated states ($I_1$ and $I_2$). $a(\varGamma)$, $b$, $c$, $d$, $e(\varGamma)$, $f$, $g$, and $h$ are the transition rates between these four states, where $a(\varGamma)$ and $e(\varGamma)$ are related to the ultrasonic radiation pressure $\varGamma$. (Bottom) Open probability of the Piezo1 channel $O_{\rm{Piezo1}}$ versus time at different ultrasonic radiation pressures $\varGamma$.}
\label{A1}
\end{figure}

\section*{APPENDIX B: Four-state model for the Piezo1 channel}
Lewis et al. express Piezo1 channels in HEK293t cells and fit the four-state gating model for the Piezo1 channel based on the data of patch clamp experiments \cite{lewis2017cellreports}. This model describes the transition of the Piezo1 channel in the HEK293t cell in four states under different pressures. In the resting state of HEK293t cells, there is a pressure of $5 \rm{mmHg}$ from the inside to the outside of the cell membrane. Here we remove the influence of the resting membrane pressure of HEK293t cells ($P_{HEK293t}$= 5 \rm{mmHg}) and use the ultrasonic radiation pressure $\varGamma$ as the pressure to control the Piezo1 channel. The schematic diagram of the modified model is shown in Fig. \ref{A1} (Top), the $\varGamma$ dependent transition rates are:
\begin{equation}
a(\varGamma)=a_0\exp(\frac{\varGamma-P_{HEK293t}}{s}), \tag{A7}
\end{equation}
\begin{equation}
e(\varGamma)=e_0\exp(\frac{P_{HEK293t}-\varGamma}{s}), \tag{A8}
\end{equation}
where transition constants $a_0= 5.1 \rm{s^{-1}}$, $b= 116.9 \rm{s^{-1}}$, $c= 8.0 \rm{s^{-1}}$, $d= 0.4 \rm{s^{-1}}$, $e_0= 34.6 \rm{s^{-1}}$, $f= 33.6 \rm{s^{-1}}$, $g= 4.0 \rm{s^{-1}}$, and $h= 0.6 \rm{s^{-1}}$. $s= 6.8 \rm{mmHg}$. The four states evolve according to the following equations:
\begin{equation}
\frac{dC}{dt}=Ob+I_1e(\varGamma)-C[a(\varGamma)+f], \tag{A9}
\end{equation}
\begin{equation}
\frac{dO}{dt}=Ca(\varGamma)+I_1d+I_2h-O(b+c+g), \tag{A10}
\end{equation}
\begin{equation}
\frac{dI_1}{dt}=Cf+Oc-I_1[e(\varGamma)+d], \tag{A11}
\end{equation}
\begin{equation}
\frac{dI_2}{dt}=Og-I_2h. \tag{A12}
\end{equation}
Figure \ref{A1} (Bottom) shows how open probabilities evolve at different ultrasonic radiation pressures.

\begin{figure}[b]
\centering
\includegraphics[width=\linewidth]{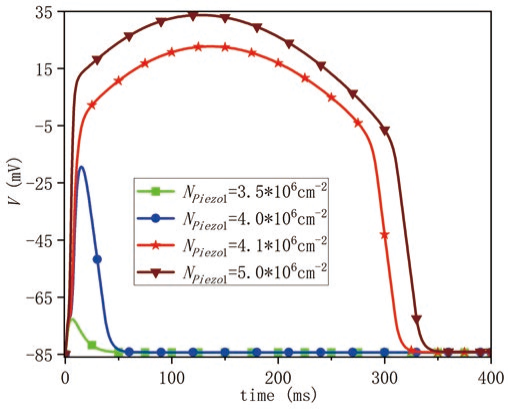}
\caption{Membrane potential $V$ of the node versus time at different Piezo1 channels' densities  $N_{\rm{Piezo1}}$. Ultrasound is applied since $t=0 \rm{ms}$, and the radiation pressure $\varGamma$ and the ultrasonic duration $D$ are $41.35 \rm{mmHg}$ and $1 \rm{ms}$, respectively. The red line shows the result at the minimum Piezo1 channels' densities that can excite the action potential. The parameters of the Fenton-Karma three-variable model are selected from the "turbulence" column in Table \ref{t1}.}
\label{A2}
\end{figure}

\section*{APPENDIX C: Modified Fenton-Karma three-variable model}
The differential equations for the three variables are as follows \cite{fenton2002chaos}:
\begin{equation}
\begin{split}
\partial_{t}V\left( {x,t} \right) = &\nabla\left( {D_V \nabla V} \right) - [I_{fi}\left( {V,v} \right) + I_{so}\left( V \right)\\
& + I_{si}\left( {V,w} \right)+ I_{Piezo1}\left( {V,\varGamma} \right)] / C_m,
\end{split}
 \tag{A13}
\end{equation}

\begin{equation}
\begin{split}
\partial_{t}v\left( {x,t} \right) = &\frac{H\left( {u_{c} - \frac{V - V_{0}}{V_{fi} - V_{0}}} \right)\left( {\text{1-}v} \right)}{H\left( {u_{v} - \frac{V - V_{0}}{V_{fi} - V_{0}}} \right)\tau_{v1}^{-} + H\left( {\frac{V - V_{0}}{V_{fi} - V_{0}} - u_{v}} \right)\tau_{v2}^{-}}\\
&-\frac{H\left( {\frac{V - V_{0}}{V_{fi} - V_{0}} - u_{c}} \right)v}{\tau_{v}^{+}}, \end{split} \tag{A14}
\end{equation}

\begin{figure}[t]
\centering
\includegraphics[width=\linewidth]{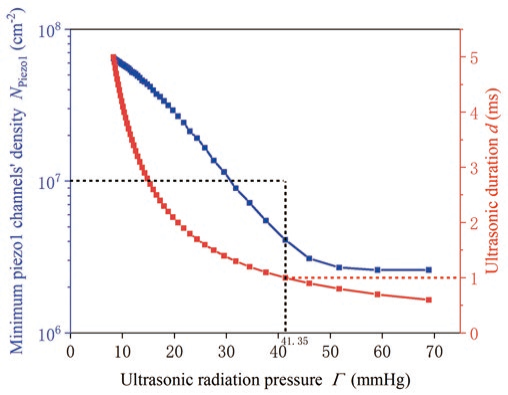}
\caption{Thresholds of Piezo1 channel's density $N_{\rm{Piezo1}}$ and the ultrasound duration $D$ at the given ultrasonic radiation pressure $\varGamma$. Blue line and circular symbols : Thresholds of $N_{\rm{Piezo1}}$ required to excite an action potential from the resting state of cardiac tissue at given $\varGamma$. Red line and square symbols: Upper limitation between $\varGamma$ and $D$ for safe sonogenetic arrhythmia control, according to the FDA guidance \cite{FDA2019}. The dashed lines indicate the values of the above parameters used in the paper.}
\label{A3}
\end{figure}

\begin{equation}
\begin{split}
\partial_{t}w\left( {x,t} \right) = &\frac{H\left( {u_{c} - \frac{V - V_{0}}{V_{fi} - V_{0}}} \right)\left( {\text{1-}w} \right)}{\tau_{w}^{-}}\\
&- \frac{H\left( {\frac{V - V_{0}}{V_{fi} - V_{0}} - u_{c}} \right)w}{\tau_{w}^{+}}, 
\end{split}
 \tag{A15}
\end{equation}
where $D_V=0.001 \rm{cm^2/ms}$ is the diffusion constant, $C_m= 1 \rm{\mu F/{cm^2}}$ is the membrane capacitance, and $H(x)$ is the standard Heaviside function. The three currents $I_{fi}$, $I_{so}$ and $I_{si}$ are given by the following equations:
\begin{equation}
\begin{split}
I_{fi}\left( {V,v} \right) = &- \frac{v}{\tau_{d}}H\left( {\frac{V - V_{0}}{V_{fi} - V_{0}} - u_{c}} \right)\left( {1 - \frac{V - V_{0}}{V_{fi} - V_{0}}} \right)\\
&\left( {\frac{V - V_{0}}{V_{fi} - V_{0}} - u_{c}} \right)[ {C_{m}\left( {V_{fi} - V_{0}} \right)}], 
\end{split}
 \tag{A16}
\end{equation}
\begin{equation}
\begin{split}
I_{so} ( V ) = & [ ( \frac{\frac{V - V_{0}}{V_{fi} - V_{0}}}{\tau_{0}} )H ( {u_{c} - \frac{V - V_{0}}{V_{fi} - V_{0}}} ) + \frac{1}{\tau_{r}}\\
& H( {\frac{V - V_{0}}{V_{fi} - V_{0}} - u_{c}} )] [{C_{m} ( {V_{fi} - V_{0}} )}], 
\end{split}
 \tag{A17}
\end{equation}
\begin{equation}
\begin{split}
I_{s i}(V, w)=&-\frac{w}{2 \tau_{s i}}\left\{1+\tanh \left[k\left(\frac{V-V_{o}}{V_{f i}-V_{o}}-u_{c}^{s i}\right)\right]\right\}\\
&\left[C_{m}\left(V_{f i}-V_{o}\right)\right], 
\end{split}
 \tag{A18}
\end{equation}
where $\tau_{v1}^{-}$, $\tau_{v2}^{-}$, $\tau_{v}^{+}$, $\tau_{w}^{-}$, $\tau_{w}^{+}$, $\tau_{d}$, $\tau_{0}$, $\tau_{r}$ and $\tau_{si}$ are the time constants, $V_{fi}$ is the reversal potential of the fast inward current, and $V_0$ is the resting potential. Different parameter sets (see Table \ref{t1}) are developed to produce the rotor and turbulence.

\begin{table}[t]
\centering
\caption{Parameter values of the Fenton-Karma three-variable model used for the simulations included in this study, modified from Fenton et al. \cite{fenton2002chaos}.}
\label{t1}
\begin{tabular}{lrrr}
Parameters & Rotor & Turbulence  \\
\midrule
$\tau_{v1}^{-}$ & 19.6 & 12\\
$\tau_{v2}^{-}$ & 1000 & 2\\
$\tau_{v}^{+}$ & 3.33 & 3.33\\
$\tau_{w}^{-}$ & 11 & 100\\
$\tau_{w}^{+}$ & 667 & 1000\\
$\tau_{d}$ & 0.41 & 0.362\\
$\tau_{0}$ & 8.3 & 5\\
$\tau_{r}$ & 50 & 33.33\\
$\tau_{si}$ & 45 & 29\\
$k$ & 10 & 15\\
$u_{c}^{si}$ & 0.85 & 0.7\\
$u_{c}$ & 0.13 & 0.13\\
$u_{v}$ & 0.055 & 0.04\\
$V_{fi}$ & 15 & 15\\
$V_{0}$ & -85 & -85\\
\bottomrule
\end{tabular}
\end{table}

\section*{APPENDIX D: SUPPLEMENTAL FIGURES AND MOVIES OF SONOGENETIC TREATMENT OF ARRHYTHMIA}
The movies are uploaded separately. Here are their descriptions. 

Movie S1. Sonogenetics for fibrillation by the whole-area ultrasound resetting of turbulence. Time $t = 0 \rm{ms}$ denotes the moment when the ultrasound is applied. Negative time images show the change in turbulence prior to the application of ultrasound.

Movie S2. Sonogenetics for fibrillation by the moving stripe-shaped ultrasound sweeping off turbulence. Time $t = 0 \rm{ms}$ denotes the moment when the ultrasound is applied. Negative time images show the change in turbulence prior to the application of ultrasound.

Movie S3. Sonogenetics for tachycardia by ultrasonic series overriding the rotor. Time $t = 0 \rm{ms}$ denotes the moment when the ultrasound is applied. Negative time images show the change in the rotor prior to the application of ultrasound.

Movie S4. Sonogenetics for tachycardia by the ultrasonic ring pulling the rotor away. Time $t = 0 \rm{ms}$ denotes the moment when the ultrasound is applied. Negative time images show the change in the rotor prior to the application of ultrasound.

Movie S5. Sonogenetics for tachycardia by ultrasonic stairs leading the rotor out. Time $t = 0 \rm{ms}$ denotes the moment when the ultrasound is applied. Negative time images show the change in the rotor prior to the application of ultrasound.

\begin{figure}[H]
\centering
\includegraphics[width=\linewidth]{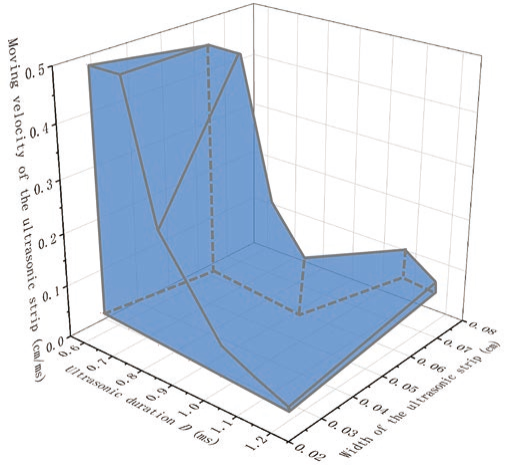}
\caption{Diagram of parameters range (blue shadow) of the ultrasonic duration $D$, the moving velocity, and the width of the ultrasonic stripe for successful and safe fibrillation termination by focused ultrasound stripe sweeping. The ultrasound radiation pressure $\varGamma$ is adjusted according to the FDA guidance about $ISPTA$.}
\label{A4}
\end{figure}

\begin{figure}[H]
\centering
\includegraphics[width=\linewidth]{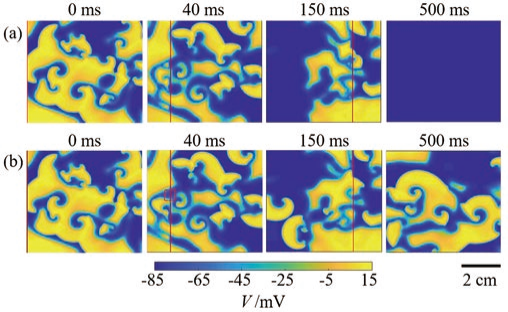}
\caption{Fibrillation termination by a focused ultrasound stripe sweeping with heterogeneous upregulation of Piezo1 channels. Subfigure (a) is the successful fibrillation termination with heterogeneity at the cell level. Subfigure (b) is the unsuccessful fibrillation termination with the tissue-level heterogeneities under the same initial condition. The red dashed box at $t=40 \rm{ms}$ indicates a region whose Piezo1 channels' density is under the threshold, not excited by a focused ultrasound stripe as in the case of cell-level heterogeneity. It leads to the failure of fibrillation termination.}
\label{A5}
\end{figure}

\begin{figure}[H]
\centering
\includegraphics[width=\linewidth]{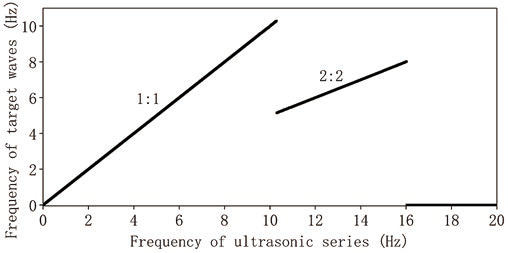}
\caption{Phase locking ratio versus the frequency of ultrasonic series. Ultrasound is applied to a circular area with a radius of $0.4 \rm{cm}$, the ultrasonic radiation pressure is $55 \rm{mmHg}$, and each ultrasonic pulse lasts for $10 \rm{ms}$. The parameters of the Fenton-Karma three-variable model are selected from the "rotor" column in Table A1. We simulate sufficient time at each frequency, removing the results of the transient state and then taking the average to obtain the frequency of the target waves.}
\label{A6}
\end{figure}

\begin{figure}[H]
\centering
\includegraphics[width=\linewidth]{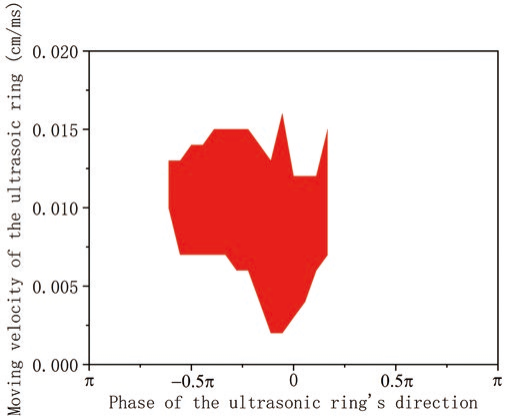}
\caption{Parameter range (red area) of the rotor can be successfully eliminated versus the ultrasonic ring's moving velocity and the phase of the ring's direction (the phase corresponding to left, down, right, and up are $\pm \pi$, $-0.5 \pi$, $0$, and $0.5 \pi$, respectively). The red area corresponds to the rotor being pulled out of the tissue boundary by the ultrasonic ring.}
\label{A7}
\end{figure}



%


\bibliography{apssamp}

\end{document}